\documentclass[aps,prl,showpacs,floatfix,twocolumn,amsmath,amssymb,preprintnumbers]{revtex4-1}
\usepackage{mathrsfs}
\usepackage[figuresright]{rotating}
\usepackage{amsmath}
\usepackage{amssymb}
\usepackage{graphicx}
\usepackage{color}
\usepackage{dcolumn}
\usepackage{bm}
\usepackage[breaklinks=true,colorlinks=true,linkcolor=blue,urlcolor=blue,citecolor=blue]{hyperref}

\makeatletter
\newcommand{\rmnum}[1]{\romannumeral #1}
\newcommand{\Rmnum}[1]{\expandafter\@slowromancap\romannumeral #1@}
\makeatother

\begin{document}

\title{Vortexlike excitations in the heavy-fermion superconductor CeIrIn$_5$}

\author{Yongkang Luo}
\email[]{ykluo@lanl.gov}
\author{P. F. S. Rosa}
\author{E. D. Bauer}
\author{J. D. Thompson}
\affiliation{Los Alamos National Laboratory, Los Alamos, New Mexico 87545, USA.}

\date{\today}

\begin{abstract}
We report a systematic study of temperature- and field-dependent charge ($\boldsymbol{\rho}$) and entropy ($\mathbf{S}$) transport in the heavy-fermion superconductor CeIrIn$_5$. Its large positive thermopower $S_{xx}$ is typical of  Ce-based Kondo lattice systems, and strong electronic correlations  play an important role in enhancing the Nernst signal $S_{xy}$. By separating the off-diagonal Peltier coefficient $\alpha_{xy}$ from $S_{xy}$, we find that $\alpha_{xy}$ becomes positive and greatly enhanced at temperatures well above the bulk $T_c$. Compared with the non-magnetic analog LaIrIn$_5$, these results suggest vortexlike excitations in a precursor state to unconventional superconductivity in CeIrIn$_5$. This study sheds new light on the similarity of heavy-fermion and cuprate superconductors and on the possibility of states not characterized by the amplitude of an order parameter.

\end{abstract}

\pacs{74.70.Tx, 74.25.fg, 74.25.Uv, 74.72.Kf}

\maketitle


Typically, a disorder-order phase transition is  described within the context of Ginzburg-Landau theory by an order parameter and identified by a spontaneously broken symmetry.
From this point of view, a superconducting transition might be special. The order parameter of superconductivity (SC) is expressed by a complex function in the form $\Psi_s(\textbf{r})$=$|\Psi_s(\textbf{r})| e^{i\theta(\textbf{r})}$\cite{GL-1950}. Gauge symmetry is broken after  phase coherence is established throughout the system. When the phase stiffness is strong, phase coherence develops \emph{concomitantly} as Cooper pairs form, and the superconducting critical temperature $T_c$ is mainly determined by $T^{MF}$, the mean-field transition temperature predicted by the BCS theory\cite{Bardeen1957a}. In contrast, if the superfluid density is small ({\it e.g.} in underdoped cuprates and organic superconductors), the phase stiffness is low, and the phase coherence can be destroyed by short-lived vortexlike excitations. In this situation, bulk SC cannot be realized until the phases of Cooper pairs are ordered, and $T^{MF}$ is simply the characteristic temperature below which pairing becomes significantly \emph{local} ($T^{MF}$$\gg$$T_c$)\cite{Emery-SCphase}. As learned from the cuprates, states without a well-defined order parameter emerge above $T_c$ and include phenomena such as superconducting phase fluctuations, pre-formed Cooper pairs, and a pseudogap.

\begin{figure}[htbp]
\vspace*{-15pt}
\hspace*{-13pt}
\includegraphics[width=9.5cm]{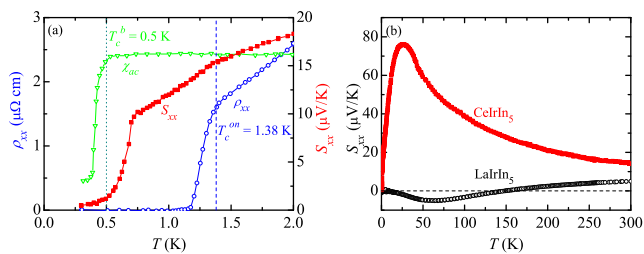}
\vspace*{-20pt}
\caption{(Color online)\label{Fig.1} (a) Temperature dependence of $\rho_{xx}$ (blue), $\chi_{ac}$ (green) and $S_{xx}$ (red) of CeIrIn$_5$, showing $T_{c}^{on}$=1.38 K and $T_c^{b}$=0.5 K. (b) Comparison of $S_{xx}(T)$ for CeIrIn$_5$ and LaIrIn$_5$.}
\end{figure}

The Ce$M$In$_5$ ($M$=Co, Rh and Ir) family of tetragonal heavy-fermion compounds is useful platform to investigate the interplay among unconventional SC, antiferromagnetic (AFM) order and spin fluctuations in the vicinity of quantum criticality. The member CeRhIn$_5$ is an incommensurate antiferromagnet at ambient pressure with N\'{e}el temperature $T_N$=$3.8$ K\cite{Hegger-CeRhIn5,Bao-CeRhIn5Neu} and can be pressurized into a superconducting state with the highest $T_c$$\sim$$2.2$ K achieved around 2.35 GPa where $T_N(p)$ extrapolates to zero\cite{Knebel-CeRhIn5Pre,Park-CeRhIn5NJP}. Textured SC was observed in the region where SC and AFM coexist, characterized by vanishingly small resistivity well above the bulk $T_c$ and the anisotropic resistive $T_c$\cite{Park-CeRhIn5Texture}, reminiscent of the nematic state observed in cuprates. In this pressure range, nuclear quadrupole resonance (NQR) experiments suggested the presence of a pseudogap that develops above $T_N(P)$ and extrapolates to the maximum in $T_c(P)$\cite{Kawasaki-CeRhIn5NQR}.
Likewise, scanning tunneling spectroscopy revealed a pseudogap that coexists with $d$-wave SC in CeCoIn$_5$\cite{Zhou-CeCoIn5STM,Wirth-Ce115STM}, and replacing a small amount of In by Cd induces coexisting AFM order and SC in CeCo(In$_{0.99}$Cd$_{0.01}$)$_5$ where again a transition to zero resistance appears well above the bulk $T_c$\cite{Park-textureSC}.
Pristine CeIrIn$_5$ shows filamentary SC\cite{Petrovic-CeIrIn5SC,Bianchi-CeIrIn5} at atmospheric pressure with a resistive onset temperature $T_c^{on}$=1.38 K, but a diamagnetic state appears only below $T_c^b$$\simeq$0.5 K [This is also illustrated in Fig.~\ref{Fig.1}(a)]. Although no direct evidence of magnetic order has yet been identified, chemical substitutions of Hg/Sn on the In site demonstrate that the SC in CeIrIn$_5$ is in proximity to an AFM quantum-critical point\cite{Shang-CeIrIn5_HgSnPt}. 
Careful magnetoresistance and Hall effect studies of CeIrIn$_5$ found evidence for a precursor state of unknown origin arising near 2 K in the limit of zero field\cite{Nair-CeIrIn52008,Nair-CeIrIn52009}. Though the pressure dependence of the precursor state is unknown, the resistive and bulk $T_c$s approach each other at the maximum in a dome of bulk SC\cite{Thompson-Ce115JPSJ}, suggesting the possibility that the precursor state may be competing with SC. The complex interplay among states in the Ce$M$In$_5$ superconductors bears strikingly similarities to the cuprates, with pure CeIrIn$_5$ at atmospheric pressure presenting an opportunity to examine more closely these similarities.

From electrical ($\boldsymbol{\rho}$) and thermoelectric ($\textbf{S}$) transport measurements in CeIrIn$_5$ and a comparison to its non-$4f$ counterpart LaIrIn$_5$, we identify signatures of vortexlike excitations well above $T_c^{on}$ ($T_c^b$). These findings suggest the existence of a pseudogaplike state where Cooper pairs start to form locally at a temperature well above $T_c^{on}$, but  phase coherence among pairs is destroyed by thermally activated vortexlike excitations, pointing to a common framework for the physics of such states in both heavy-fermion and cuprate\cite{Scalapino-RMP2012}.


Single crystalline CeIrIn$_5$ was grown from an indium flux method\cite{Petrovic-CeIrIn5SC}. The crystal was pre-screened by both resistivity and magnetic susceptibility measurements to ensure the absence of free In. Thermoelectric measurements were carried out by means of a steady-state technique. A pair of well calibrated differential Chromel-Au$_{99.93\%}$Fe$_{0.07\%}$ thermocouples was used to measure the temperature gradient. Upon a thermal gradient $-$$\nabla T$$\parallel$$\textbf{x}$ and a magnetic field $\textbf{B}$$\parallel$$\textbf{z}$, both thermopower signal $S_{xx}$=$-E_x/|\nabla T|$ and Nernst signal $S_{xy}$=$E_y/|\nabla T|$ were collected by scanning field at fixed temperatures. The same contact geometry also was used to measure electrical resistivity ($\rho_{xx}$) and Hall resistivity ($\rho_{yx}$). Both electrical and thermal currents were applied along the $\textbf{a}$-axis, and the magnetic field was parallel to $\textbf{c}$. The same measurements were  performed on the non-magnetic analog LaIrIn$_5$ for comparison. We adopt the sign convention as Ref.~\cite{Bridgman-Sign}, which defines a \emph{positive} Nernst signal for vortex motion\cite{Xu-LSCONernst,Wang2006}.


In the presence of a temperature gradient $-$$\nabla T$, an electric field $\textbf{E}$ and a magnetic field $\textbf{B}$, the total current density is $\textbf{J}$=$\boldsymbol{\sigma}$$\cdot$$\textbf{E}$+$\boldsymbol{\alpha}$$\cdot$$(-\nabla T)$, where $\boldsymbol{\sigma}$ is the conductivity tensor, and $\boldsymbol{\alpha}$=$\frac{\pi^2k_B^2T}{3q}\frac{\partial\boldsymbol{\sigma}}{\partial\varepsilon}|_{\varepsilon=\varepsilon_F}$ ($k_B$ is Boltzman constant, $q$ is charge of carriers, $\varepsilon_F$ is chemical potential) is the Peltier conductivity tensor\cite{Ziman}. In an equilibrium state without net current, the Boltzman-Mott transport equation deduces the thermoelectric tensor
\begin{equation}
\textbf{S}=\boldsymbol{\alpha}\cdot\boldsymbol{\sigma}^{-1}=\boldsymbol{\alpha}\cdot\boldsymbol{\rho}.
\label{Eq.1}
\end{equation}

We start with the temperature dependence of thermopower $S_{xx}(T)$ as shown in Fig.~\ref{Fig.1}(b). $S_{xx}(T)$ of LaIrIn$_5$ is positive at room temperature and changes sign near 150 K, characteristic of the expected multi-band behavior\cite{Shishido-Ce115FS}.
In contrast, $S_{xx}(T)$ of CeIrIn$_5$ is positive in the full temperature range between 0.3 K and 300 K, displaying a pronounced maximum at around 25 K with the magnitude reaching 76 $\mu$V/K. This peak in $S_{xx}(T)$ is associated with the onset of Kondo coherence\cite{Takaesu-CeIrIn5Pre}. These features are consistent with a Ce-based Kondo lattice in which the strong hybridization between $4f$- and conduction-electrons forms a Kondo resonance with the density of states $N({\varepsilon})$ asymmetric with respect to $\varepsilon_F$\cite{Zlatic-CeYbS,Miyake-Seebeck2005} (see below). At low temperatures, $S_{xx}(T)$ shows a small kink at $T_c^{on}$=1.38 K, but  drops sharply at 0.7 K and tends to saturate below $T_c^b$=0.5 K [cf Fig.~\ref{Fig.1}(a)]. Down to the lowest temperature of 0.3 K, however, $S_{xx}(T)$ still remains finite. We attribute this non-vanishing $S_{xx}$ in the bulk superconducting state  to the low $T_c^b$ of CeIrIn$_5$: even a small temperature gradient may generate ungapped quasiparticles that contribute transport entropy.

\begin{figure}[htbp]
\vspace*{-15pt}
\hspace*{-13pt}
\includegraphics[width=9.5cm]{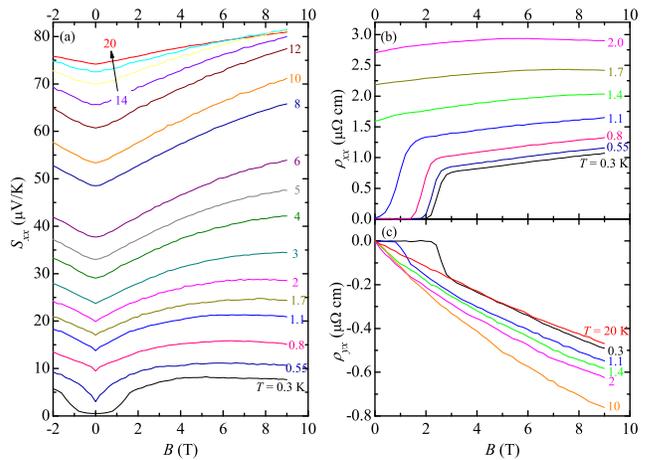}
\vspace*{-20pt}
\caption{(Color online)\label{Fig.2} (a) Field dependence of $S_{xx}$ of CeIrIn$_5$ at selected temperatures. (b) and (c) display $\rho_{xx}(B)$ and $\rho_{yx}(B)$, respectively. }
\end{figure}

Figure~\ref{Fig.2}(a) displays isothermal field dependence of $S_{xx}$ at various temperatures. For all temperatures, the magneto-thermopower is positive. One important feature of $S_{xx}(B)$ is a valley in the vicinity of zero field. As temperature decreases, this valley deepens and evolves into a cusp when $T$$\leq$3 K. At 0.3 K, $S_{xx}$ is small at $B$=0 but recovers when the field is larger than 1.6 T. With the field dependencies of $\rho_{xx}$ and $\rho_{yx}$ shown in Fig.~\ref{Fig.2}(b) and (c),  respectively, it is reasonable to attribute this small transport-entropy state to a SC state. The cusp in $S_{xx}(B)$ occurring near 3 K is indicative of the loss of transport entropy well above $T_{c}^b$. The critical field recovering a normal state, however, is much smaller than that determined from $\rho_{xx}(B)$ [Fig.~\ref{Fig.2}(b)] and $\rho_{yx}(B)$ [Fig.~\ref{Fig.2}(c)]. Systematic analysis of $\rho_{xx}(B)$ and $\rho_{yx}(B)$ by Nair {\it et al.}\cite{Nair-CeIrIn52008,Nair-CeIrIn52009} showed that the modified Kohler's scaling [$\Delta\rho_{xx}(B)/\rho_{xx}(0)$$\propto$$\tan^2\theta_H$, where $\theta_H$=$\arctan(\rho_{yx}/\rho_{xx})$ is the Hall angle] breaks down prior to $T_c^{on}$, the region where we observe a large Nernst effect (see below). Similar phenomenon was observed in CeCoIn$_5$ and CeRhIn$_5$ under pressure\cite{Nakajima-CeMIn5JPSJ}, as in cuprates, and is reminiscent of a pseudogaplike precursor state\cite{Abe-YBCOHall}.

\begin{figure}[htbp]
\vspace*{-30pt}
\includegraphics[width=7.5cm]{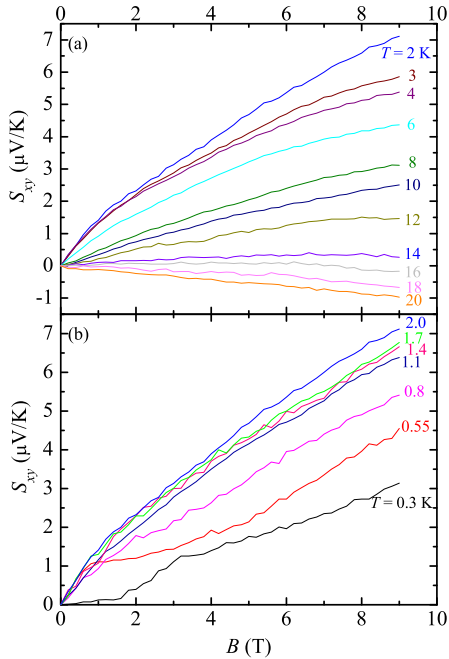}
\vspace*{-20pt}
\caption{(Color online)\label{Fig.3} Nernst signal $S_{xy}$ of CeIrIn$_5$ as a function of $B$ at selected temperatures. (a), $0.3$$\leq$$T$$\leq$$2.0$ K; (b), $2$$\leq$$T$$\leq$$20$ K. }
\end{figure}

In Fig.~\ref{Fig.3} we present the field dependence of the Nernst signal $S_{xy}$, the off-diagonal term of the thermoelectric tensor $\textbf{S}$. $S_{xy}(B)$ is both negative and linear in $B$ at 20 K. The magnitude of $S_{xy}(B)$ decreases with decreasing $T$ and changes sign near 15 K [Fig.~\ref{Fig.3}(a)]. The non-linearity of $S_{xy}(B)$ becomes pronounced and the value of $S_{xy}$ rapidly increases with decreasing $T$. At 2 K, $S_{xy}$ reaches 7 $\mu$V/K when $B$ is 9 T. We will see that such a large $S_{xy}$, even larger than that in the vortex-liquid state of cuprates\cite{Xu-LSCONernst,Wang2006}, is mainly due to the Kondo effect, albeit the vortexlike excitation contribution is also non-negligible. A large Nernst effect also has been seen in other Kondo-lattice compounds, like CeCoIn$_5$\cite{Bel-CeCoIn5Nernst,Onose-CeCoIn5Nernst,Izawa-CeCoIn5Nernst}, CeCu$_2$Si$_2$\cite{SunP-CeCu2Si2Nernst}, URu$_2$Si$_2$\cite{Yamashita-URu2Si2Nernst} and SmB$_6$\cite{LuoY-SmB6Thermo}. In CeIrIn$_5$ $S_{xy}$ starts to drop when $T$ is lower than 2 K but remains positive down to 0.3 K, the base temperature of our measurements [Fig.~\ref{Fig.3}(b)]. At 0.3 K, which is below $T_c^b$, $S_{xy}(B)$ increases slowly at small field but much more rapidly near 1.8 T. It is likely that this 1.8 T magnetic field defines a melting field $B_m$ above which the vortex solid melts into a vortex-liquid state. A large number of vortices start to move in response to a temperature gradient and this results in the abrupt increase in $S_{xy}(B)$. Similar results also have been seen in other type-\Rmnum{2} superconductors, like cuprates\cite{Xu-LSCONernst,Wang2006} and CeCoIn$_5$\cite{Onose-CeCoIn5Nernst}. This vortex-lattice melting field disappears immediately when $T$ exceeds $T_c^b$, {\it e.g.} 0.55 K as shown in Fig.~\ref{Fig.3}(b). This implies that a well-defined Abrikosov- lattice of vortices only exists in the bulk superconducting state of CeIrIn$_5$.

\begin{figure}[htbp]
\vspace*{-15pt}
\hspace*{-5pt}
\includegraphics[width=9cm]{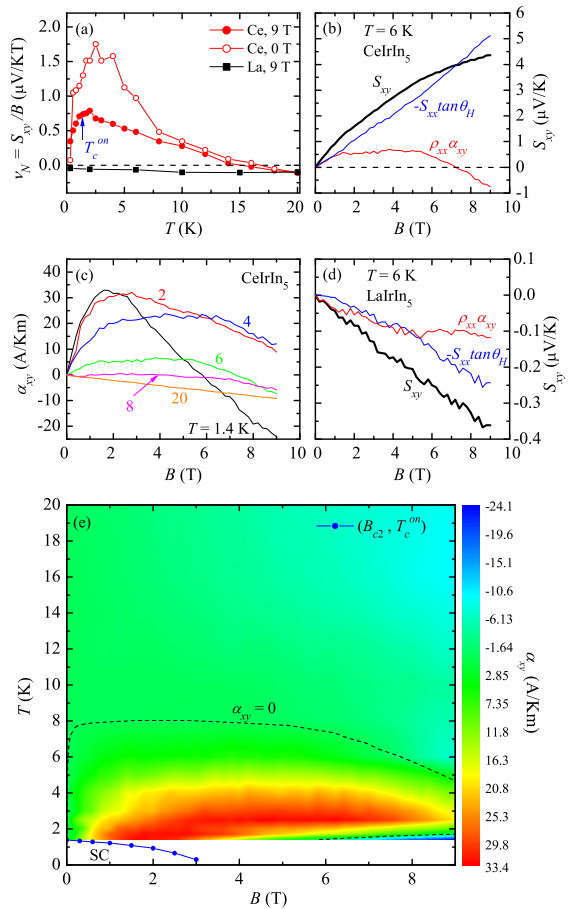}
\vspace*{-25pt}
\caption{(Color online)\label{Fig.4} (a) Temperature-dependent Nernst coefficient $\nu_N$ of LaIrIn$_5$ and CeIrIn$_5$. For CeIrIn$_5$, the open symbols are the initial slopes of $S_{xy}(B)$ as $B$$\rightarrow$0. (b) and (d) show the separation of $\rho_{xx}\alpha_{xy}$ from $S_{xy}$ at $T$=6 K for CeIrIn$_5$ and LaIrIn$_5$, respectively. (c) Off-diagonal Peltier coefficient $\alpha_{xy}$ as a function of $B$ at selected temperatures. (e) Contour plot of $\alpha_{xy}(B, T)$, with the resistively determined $B_{c2}(T)$ shown in the lower left corner. The black dash line is the boundary where $\alpha_{xy}$=0.}
\end{figure}

Figure~\ref{Fig.4}(a) shows the temperature dependence of the Nernst coefficient $\nu_N$$\equiv$$S_{xy}/B$. Here, the solid symbols are obtained at $B$=9 T, and the open symbols represent the initial slope of $S_{xy}(B)$ as $B$$\rightarrow$0. In both definitions, $\nu_N$ above $T_c^{on}$ is large and sign-changes near 15 K. It is well known that for a single-band, non-superconducting and non-magnetic metal, the Nernst signal is vanishingly small, due to so-called Sondheimer cancellation\cite{Sondheimer},
\begin{equation}
S_{xy}=\rho_{xx}\alpha_{xy}-S_{xx}\tan\theta_H.
\label{Eq.2}
\end{equation}
A large Nernst effect has been observed in: (\rmnum{1}) multi-band systems such as NbSe$_2$\cite{Bel-NbSe2Nernst} in which the ambipolar effect violates Sondheimer cancellation; (\rmnum{2}) phase slip due to vortex motion in type-\Rmnum{2} superconductors, as in underdoped cuprates\cite{Xu-LSCONernst,Wang2006}; (\rmnum{3}) ferromagnets like CuCr$_2$Se$_{4-x}$Br$_x$ in which $S_{xy}(B)$ scales to magnetization $M(B)$, known as anomalous Nernst effect\cite{Lee-CuCr2Se4ANE}; (\rmnum{4}) Kondo-lattice systems, like CeCu$_2$Si$_2$, in which an enhanced $\nu_N$ is determined by  asymmetry of the on-site Kondo scattering rate\cite{SunP-CeCu2Si2Nernst}.

We can exclude the anomalous Nernst effect in CeIrIn$_5$ because $S_{xy}(B)$ does not scale with the magnetization, which is essentially a linear function of $B$ (data not shown). From the negative Hall resistivity $\rho_{yx}(B)$ shown in Fig.~\ref{Fig.2}(c), we also rule out a substantive contribution from skew scattering because, as discussed in Refs.~\cite{Nakajima-CeMIn5JPSJ,LuoY-CeNi2As2Pre}, it generates a \emph{positive} anomalous Hall effect for Ce ions.

To study a possible multiband contribution to the Nernst signal of CeIrIn$_5$, we performed the same measurements on the non-$4f$ counterpart LaIrIn$_5$. According to quantum oscillation measurements and density functional theory (DFT) calculations, LaIrIn$_5$ is electron-hole compensated \cite{Shishido-Ce115FS,Hall-LaTIn5dHvA}, and a large Nernst effect is possible\cite{Bel-NbSe2Nernst}. The Nernst signal of LaIrIn$_5$, however, is surprisingly both negative and linear in $B$ [see Fig.~\ref{Fig.4}(d) for instance], and most importantly, the Nernst coefficient remains small between 0.3 K and 20 K [Fig.~\ref{Fig.4}(a)]. This demonstrates that a multiband effect does not play an important role in LaIrIn$_5$. Compared with LaIrIn$_5$, CeIrIn$_5$ has a somewhat larger Fermi surface due to a partially itinerant $4f$-band\cite{Shishido-Ce115FS}, electron-hole compensation is relatively unbalanced, and, therefore, a multiband contribution to the Nernst signal of CeIrIn$_5$ is expected to be even weaker.

To better understand the origin of a large Nernst effect in CeIrIn$_5$, we separate  $\rho_{xx}\alpha_{xy}$ from the total Nernst signal $S_{xy}$ [cf Eq.~(\ref{Eq.2})]. As an example, we show $S_{xy}$, $\rho_{xx}\alpha_{xy}$ as well as $-S_{xx}\tan\theta_H$ at 6 K in Fig.~\ref{Fig.4}(b). As seen, $-S_{xx}\tan\theta_H$ is the dominant contribution to $S_{xy}$. In a Kondo-lattice system, strong electronic correlations build up a resonance in the density of states near the chemical potential $\varepsilon_F$, and the scattering rate ($1/\tau$) is now mainly determined by the very narrow, renormalized $4f$-bands, {\it i.e.} $N_f(\varepsilon)$. As a result, the thermopower, given by Eq.~(\ref{Eq.3}), becomes large\cite{Kamran-Thermopower}
\begin{equation}
S_{xx}\propto\frac{\partial\ln\tau}{\partial\varepsilon}\propto-\frac{\partial\ln N_f(\varepsilon)}{\partial\varepsilon}|_{\varepsilon=\varepsilon_F}
\label{Eq.3}
\end{equation}
due to an asymmetric $N_f(\varepsilon)$ and is reflected in data plotted in Fig.~\ref{Fig.1}(b). This asymmetry of on-site Kondo scattering also enters $S_{xy}$ through the term $-S_{xx}\tan\theta_H$ and gives rise to the large Nernst effect in CeIrIn$_5$ and other Kondo-lattice systems as well\cite{Bel-CeCoIn5Nernst,Onose-CeCoIn5Nernst,SunP-CeCu2Si2Nernst,LuoY-SmB6Thermo}.

We note that $-S_{xx}\tan\theta_H$ surpasses $S_{xy}$ when $B$ is larger than 7.3 T at 6 K, and this leads to a sign change in $\rho_{xx}\alpha_{xy}$ [Fig.~\ref{Fig.4}(b)]. Figure ~\ref{Fig.4}(c) shows the field dependent $\alpha_{xy}$ at various temperatures. Due to a large contribution from asymmetric Kondo scattering in $S_{xy}(B)$, $\alpha_{xy}(B)$ clearly differs from $S_{xy}(B)$ and, therefore, more intrinsically describes the off-diagonal thermoelectric response. $\alpha_{xy}(B)$ is negative and linear in $B$ at 20 K. As $T$ decreases, an anomalous positive term gradually appears on top of the negative linear background. Similar behavior was observed in CeCoIn$_5$ and was interpreted as a signature of phase-slip events caused by the passage of individual vortices\cite{Onose-CeCoIn5Nernst}. To compare, we show $\rho_{xx}\alpha_{xy}$ at 6 K for LaIrIn$_5$ in Fig.~\ref{Fig.4}(d). As expected, the unusual behavior is absent in LaIrIn$_5$ where there is only a small negative $\rho_{xx}\alpha_{xy}$.

It is reasonable to write $\alpha_{xy}$ in the form\cite{Onose-CeCoIn5Nernst}
\begin{equation}
\alpha_{xy}=\alpha_{xy}^n+\alpha_{xy}^s,
\label{Eq.4}
\end{equation}
where $\alpha_{xy}^n$ is the contribution from normal quasiparticles and $\alpha_{xy}^s$ represents an anomalous term stemming from vortex excitations. The positive $\alpha_{xy}(B)$ manifests that vortex motion dominates the quasiparticle term. We summarize these results in a contour plot of $\alpha_{xy}(B,~T)$ in Fig.~\ref{Fig.4}(e). Below the $\alpha_{xy}$=0 boundary near 8 K, vortexlike excitations contribute and become most pronounced in the ``island" region below 4 K. These temperature scales are qualitatively different from those in CeCoIn$_5$ in which Nernst effect develops at very low temperature near a field-induced quantum-critical point\cite{Izawa-CeCoIn5Nernst}. We also note that the temperature dependence of $\alpha^s_{xy}/B$ in CeIrIn$_5$ cannot be reproduced even approximately by assuming that it arises from Gaussian superconducting fluctuations (data not shown) which seems successful in describing the Nernst effect for optimally-doped and overdoped cuprates but not underdoped ones\cite{Ussishkin-Gaussian}. Taking $T_c^{on}$=1.38 K in simulation, the calculated $\alpha_{xy}^s/B$ by Gaussian model is an order of magnitude smaller than the observed values. These findings suggest that local Cooper pairs start to form at a temperature well above $T_c^{on}$ and that phase coherence among them is destroyed by thermally activated vortexlike excitations. We estimate the phase-order temperature (above which the phase coherence is destroyed), $T_{\theta}^{max}$$\sim$4 K, if we adopt Emery's model\cite{Emery-SCphase} to CeIrIn$_5$ with  lattice parameter $c$=7.515 \AA\cite{Petrovic-CeIrIn5SC} and superconducting penetration depth $\lambda(0)$$\sim$10$^4$ \AA\cite{Vandervelde-CeIrIn5Lambda}. The ratio $T_{\theta}^{max}/T_{c}^b$$\sim$8 (or $T_{\theta}^{max}/T_{c}^{on}$$\sim$2.9) is significantly smaller than that of conventional superconductors (10$^2$$\sim$10$^5$) but is comparable to that of underdoped high-$T_c$ cuprates ($<$10)\cite{Emery-SCphase} whose phase stiffness is soft. Perhaps not coincidentally, $T_{\theta}^{max}$ is comparable to the estimated zero-field temperature of a precursor state found in magnetotransport \cite{Nair-CeIrIn52008,Nair-CeIrIn52009}. The filamentary nature of SC\cite{Bianchi-CeIrIn5} also would imply a dilute superfluid density, which renders the phase fluctuations possible in CeIrIn$_5$\cite{Emery-SCphase}. Finally, we note that the specific heat ($C/T$) of CeIrIn$_5$ deviates from a $-\log T$ dependence below $\sim$2-4 K where it rolls over to a weaker (nearly constant) temperature dependence\cite{Borth-CeIrIn5C}. On a similar temperature scale, $^{115}$In nuclear spin-lattice relaxation rate ($1/{T_1}$) also shows a weak inflection at around 6 K\cite{ZhengG-CeIrIn5NQR}. These evolutions prior to $T_c$ suggest formation of a partial gap in $N({\varepsilon})$ that is in parallel with ungapped heavy quasiparticles. Whether these behaviors are the consequences of a possible pseudogap or correlated with the formation of local Cooper pairs is still an open question and requires further investigation.


Thermoelectric measurements in combination with charge transport in the heavy-fermion superconductor CeIrIn$_5$ indicate the formation  of an unusual state above $T_c$ that is reminiscent of cuprate physics. By separating the off-diagonal Peltier coefficient $\alpha_{xy}$ from $S_{xy}$, we find that $\alpha_{xy}$ becomes positive and  greatly enhanced at the temperatures well above $T_c$. Compared with the non-magnetic analog LaIrIn$_5$, these results suggest vortexlike excitations in a precursor state of CeIrIn$_5$. This  work sheds new light on bridging the similarity between heavy-fermion and cuprate superconductors and is a step towards uncovering the mechanism of the unconventional superconductivity in the Ce$M$In$_5$ family compounds.


We thank Shizeng Lin for insightful conversations. Work at Los Alamos was performed under the auspices of the U.S. Department of Energy, Division of Materials Sciences and Engineering. P. F. S. Rosa acknowledges support through a Director's Postdoctoral Fellowship that is funded by the Los Alamos LDRD program.


%

\end{document}